\documentclass[preprint2]{aastex}
\slugcomment{submitted to {\it The Astronomical Journal}}
\begin{document}
\title{$Caby$ Photometry of the Hyades: Comparisons to the Field Stars}
\author{Barbara J. Anthony-Twarog, Bruce A. Twarog, and Judy Yu}
\affil{Department of Physics and Astronomy, University of Kansas, 
Lawrence, KS 66045-2151}
\affil{Electronic mail: bjat@ukans.edu,twarog@ukans.edu,judyyu@ku.edu}
\begin{abstract}
Intermediate-band photometry of the Hyades cluster on the {\it Caby} system is
presented for dwarf stars ranging from spectral type A through late K. A mean
$hk$, $b-y$ relation is constructed using only single stars without 
anomalous atmospheres and compared to the field stars of the solar
neighborhood. For the F dwarfs, the Hyades relation defines an approximate
{\it lower} bound in the two-color diagram, consistent with an [Fe/H] between
+0.10 and +0.15.  These index-color diagrams follow the common convention of
presenting stars with highest abundance at the bottom of the plot although
the index values for the metal-rich stars are numerically larger.
For field F dwarfs in the range [Fe/H] between +0.4 and --1.0, 
[Fe/H] = --5.6 $\delta hk$ + 0.125, with no evidence
for a color dependence in the slope. For the G and K dwarfs, the Hyades mean 
relation crosses the field star distribution in the two-color diagram, 
defining an approximate $upper$ bound for the local disk stars. Stars
found above the Hyades stars fall in at least one of three categories:
[Fe/H] below --0.7, 
[Fe/H] above that of the Hyades, or chromospherically active. It is concluded 
that, contrary to the predictions of model atmospheres, the $hk$ index for 
cool dwarfs at a given color hits a maximum value for stars below solar 
composition and, with increasing [Fe/H] above some critical value, declines. 
This trend is consistent, however, with the predictions from synthetic indices 
based upon much narrower Ca filters where the crossover is caused by 
the metallicity sensitivity of $b-y$. 
 
\end{abstract}
\keywords{open clusters:individual (Hyades)---stars:abundances---techniques:photometric}
\section{INTRODUCTION}
Among well-studied open clusters, the Hyades occupies a unique 
niche due to the combination of its proximity and its high metallicity
\citep{TAT97}.
This rich sample of nearby stars of common age and composition provides
a matchless testbed for any investigation of purely temperature-dependent
trends at a given [Fe/H] among dwarfs with a wide range of mass. In
particular, the high metallicity of the Hyades has led to its adoption as
a reference point for zeroing and/or testing the metallicity scale for
several photometric systems, including the {\it UBV} \citep{CA79},
the DDO \citep{DEM77,TAT96}, and the Str\"omgren \citep{CR75, SN89} systems.
The purpose of this investigation is
to add another, the {\it Caby} system, to the long list of
fundamental color relations defined by the Hyades, while
investigating the effects of metallicity, and potentially of age, on the $hk$
index for a large sample of nearby field stars.

The {\it Caby} system represents an extension of the traditional four-color,
$uvby$ intermediate-band photometric system to a fifth filter centered on the
H and K lines of Ca II. Details of the filter definition and design as well as 
the fundamental standards may
be found in \citet{ATT91}, while an extensive catalog of stars observed on the
system and tied to the $b-y$ scale of \citet{OLS93} may be found in 
\citet{TAT95}. The filter was designed initially with metal-deficient stars 
in mind, as demonstrated by numerous applications to date on normal field stars 
\citep{ATT91, AST92, ATT98, ATT00}, clusters \citep{ATC95, RE00}, and variables
\citep{BD96, HI98}.
Metallicity calibrations have been produced for both the metal-deficient giants
\citep{ATT98} and metal-deficient 
dwarfs \citep{ATT00}, but preliminary analysis indicated that
for dwarfs hotter than the sun, the $hk$ index, defined as {\it (Ca-b)-(b-y)},
remains metallicity sensitive for stars of solar abundance or higher 
\citep{TAT95}, a result consistent with the theoretical models of \citet{SO93}.
Because of the high metallicity of the Hyades relative to the typical star in
the field of the solar neighborhood, it provides an ideal test of this
prediction, as well as a means of probing the limits of its sensitivity. 

A second issue of particular relevance for metal-rich stars is the role of
chromospheric emission on the apparent strength of the $hk$ index. As one
moves toward lower temperature, the expectation is that the absorption features
will saturate, leaving an index which is almost exclusively dependent upon 
temperature. That the sensitivity of the {\it hk} index to metallicity
should decline near [Fe/H] = 0.0 for $(b-y)$ redder than 0.5 is readily 
apparent in the synthetic indices
of \citet{SO93}, though the models imply that even G and K dwarfs with
abundances well above solar should exhibit larger $hk$ indices at a given
color. However, line reversals triggered by chromospheric activity
\citep[see, e.g.,][]{VA80}
may fill in the cores of the absorption features, making the star appear more
metal-deficient than it actually is. Moreover, the strength of this line 
reversal has long been known to be age-dependent \citep{WI63}. 
Since the Hyades is moderately young (less than 1 Gyr) compared to the 
average cool dwarf near the sun and
contains some stars with an anomalous degree of chromospheric activity, it may
provide some insight into this question. Finally, the majority
of the stars known to be Hyades members have been studied in detail,
thereby allowing us to look for additional photometric anomalies not tied to 
the metallicity of the stars.

Section 2 contains the details of the observations, their transformation
to the standard $Caby$ system, and the merger of our
$V$, $b-y$ data with an extensive
array of published data on the Hyades dwarfs in an effort to minimize
the potential effects of internal photometric scatter on the mean relations. In
Sec. 3 we derive the Hyades mean relation in the {\it hk, (b-y)} diagram and
discuss the potential sources of intrinsic scatter in the two-color
diagram. Sec. 4 compares the single-star Hyades relation to a large
sample of nearby field stars, providing some insight into possible sources
of the differences between expectation and reality, particularly at cooler
temperatures. Sec. 5
contains a summary of our conclusions and suggestions for further work on the
system in light of the Hyades anomalies. 

\section{The Data}

\subsection{Observations}
The data discussed here were obtained as part of the wide-ranging
program of observations on the {\it Caby} system started in Dec. 1983 and
continued on through 1994. The {\it Caby} work since then has focused 
on more specialized
applications, but some Hyades stars were included from the beginning of the
program with calibration and standardization purposes in mind. The 
preliminary sample was selected from the well-known, brighter members 
identified by \citet{VB52}, but the more comprehensive database detailed
here is tied to the compilation of \citet{GR88}. The observations
were made on a variety of telescopes including the 0.9 m and 1.2 m at KPNO
and, primarily, the 0.6 m, 0.9 m, 1.0 m, and 1.5 m at CTIO. All the telescopes
at the national observatories were equipped with pulse-counting, single-channel
photometers mounted with S-20 photomultipliers. In addition, a modest number
of observations were obtained using a pulse-counting photometer equipped 
with a 1P21 photomultiplier on the 0.4 m telescope at Braeside Observatory in 
Arizona. The filter sets used have changed over the years, but extreme care
has been taken to ensure transformation to the standard system and no
significant problems have been encountered. Observations of the fainter stars
in this study were all made with the 1.0 m and 1.5 m telescopes at CTIO
equipped with the same filter set. For more information regarding 
the design of the $Ca$ filter and the observing procedure, the 
reader is referred to \citet{ATT91}.

\subsection{Reduction and Transformation}
A complete description of the methodology used to merge all the observations
taken from the multiple runs over a number of years is given in \citet{TAT95}
and will not be repeated here. A key point is that the definition of the
standard $b-y$ system was changed between the publication of the primary
standards \citep{ATT91} and the final catalog. This was prompted by the
extension of the system to include a significant fraction of cool dwarfs and
the publication of the definitive set of cool star observations on the
$uvby$ system by \citet{OLS93}. The composite, homogenized sample on the 
instrumental system included all of our Hyades data up to that time; this
sample was used to define the transformation in $b-y$, with separate 
transformation curves defined for hotter stars, cool dwarfs, and cool giants. 
In the mean, the large overlap between the $Caby$ and $uvby$ catalogs produced 
an excellent match between the two systems. Though several brighter stars 
in the Hyades were included in the 1995 catalog, the fainter dwarfs were 
not due to the optimistic expectation of additional observations. 

We have accumulated over 535 observations on the system of the
$Caby$ catalog \citep{TAT95} for 114 Hyades dwarfs.  As the $b-y$
colors of that catalog were intended to be consistent with \citet{OLS93},
the Hyades data should be also; indeed, a few of the brighter
Hyades stars were included
in the analysis from which the color transformations to \cite{OLS93} were
developed.  Yet, it is not uncommon for cluster photometric samples to show 
small offsets with respect to a standard system even though the all-sky
data reveal no systematic differences.  We were particularly concerned because
most of the Hyades stars were observed at CTIO at large airmass. To minimize
the discrepancies with the \cite{OLS93} system, we directly compared
the Hyades stars common to both catalogs; Hyades stars
from \citet{OLS94} were also included in this comparison.

The results of the comparison revealed slight differences in our colors and
magnitudes with respect to \citet{OLS93}.  For 81 stars stars bluer than
$b-y = 0.50$, our preliminary colors were too red by 
$0.0049 \pm 0.0044$ (s.d.).  The $V$ magnitudes are
correspondingly brighter by $0.006 \pm 0.018$ (s.d.).
For the 26 redder stars, a linear relation between the $b-y$ colors was
derived; the transformation between our preliminary colors
and those of \citet{OLS93} was  $(b-y)_{OLS} = 
1.049 (b-y)_{ATT} - 0.0264$. The points scatter about the mean relation
with residuals amounting to only $\pm 0.0055$ mag (s.d.).  A similar 
adjustment was required for our preliminary $V$ magnitudes, i.e.,
$V_{OLS} = V_{ATT} + 0.1 (b-y)_{ATT} - 0.0391$ with a scatter of $\pm 0.0086$.  
The new Hyades $Caby$ data presented in Table 1 (columns 4 through 9) have been 
adjusted from our preliminary results according to these precepts
and are solidly tied to the \citet{OLS93} system.  The identification
number found in the WEBDA database is found in column 1, with HIC, HD 
or other identifications listed in column 2.  Notes in column 3 refer
to information gathered about suspected binaries and non-members, described
further in Sec. 3.

In an effort to further improve the final precision of the Hyades $b-y$ 
colors and magnitudes, we searched the internet open cluster database, WEBDA, to
find Str\"omgren photometric datasets in the Hyades which substantially 
overlapped with our Hyades data.  Published $V, b-y$ data from 
\citet{OLS83, OLS93, OLS94}, \citet{rf92}, and \citet{sw93} were used.  
Most of the other sizable Hyades Str\"omgren photometric samples are 
restricted to brighter and hotter main sequence stars and were not 
included in the merger.  Of the datasets included in the merged sample,
that of \citet{cp66} does not include $V$ magnitudes.  
One large sample, that of \citet{p69} provided only a rather noisy 
comparison to our $(b-y)$ colors and was not used. Finally, we 
also included unpublished data for 10 Hyades dwarfs obtained by 
Bruce Carney with the original prototype of the $Ca$ filter; only 
$(b-y)$ colors and $hk$ indices were available.  

As expected, small offsets between these surveys and our preliminary 
sample were found, though with no evidence for color terms. Typical 
differences in $b-y$ ranged from --0.005 to +0.015 mag, with a range of
$\pm 0.016$ for average differences in $V$ magnitudes.  The indices
from each dataset were adjusted for these small differences, placing
them on our preliminary system, then transformed
to the \citet{OLS93} system according to the precepts indicated above.
The data sets were finally merged and averaged using weights based 
upon the consistency between each dataset and our preliminary Hyades dataset.  
The final averaged $V$ magnitudes and $b-y$
colors are found in columns 11 and 12 of Table 1, followed by the
number of $V$ and $b-y$ observations incorporated in the average and
the number of datasets included.  The standard errors of the mean for the
final magnitudes and indices were quite small and are not included in Table 1.
Typical values of $\pm 0.004$ (s.e.m.) and
$\pm 0.003$ (s.e.m.) were found for $V$ and $b-y$, respectively.

Finally, the $b-y$ colors for Hyades stars in common with
\citet{OLS93} were checked to verify that the color system is consistent with
that sample.  From 81 stars in common, the mean difference in $(b-y)$ for 
the merged sample is
$-0.0010 \pm 0.0037$ (s.d.) in the sense (OLS - Table 1).

\section{The Hyades $hk - (b-y)$ Relation}
           
In defining any two-color relation, one ideally wants to exclude stars
affected by anomalies which might distort the relation beyond normal 
photometric scatter. In the case of a cluster sample, a fundamental 
requirement is that stars be classified
as cluster members. Since the time of the original compilation of the
Hyades stars for observation, based primarily upon the discussion in 
\citet{GR88}, additional parallax, radial-velocity, and proper-motion
observations, particularly those by 
$Hipparcos$ \citep{PE98}, have clarified and eliminated
the membership of a few stars in Table 1.  Star VB/WEBDA 9, despite positions
in the color-magnitude diagram (CMD) and $hk, b-y$ diagram consistent with 
the Hyades, is now classed as a kinematic non-member. A second, more obvious
interloper is VB/WEBDA 172. Originally classed as a Hyades subdwarf, it lies
well below the main sequence in the Hyades CMD. This class of stars is now
known to be a product of more distant nonmembers whose space velocities
conspire to produce proper motions compatible with those of the Hyades members
\citep{HA81}. These two stars will be removed from further discussion. The
remaining 112 stars are classed as probable members by \citet{PE98} or,
if not included in the $Hipparcos$ survey, have been tagged as probable
members by \citet{GR88}.

The next obvious exclusion to make is the elimination of all stars which are
multiple systems. The presence of one or more companions may distort the
colors of the composite system directly via the combined colors of the stars
or indirectly through an alteration of the structure and/or evolutionary 
state of the star and its atmosphere, as in the case of chromospherically 
active stars created through tidal interaction. Following \citet{dB01}, 
in deriving the mean relation we have eliminated any star 
which has been tagged as a potential binary through direct observation from the
ground or from data obtained from $Hipparcos$; see Table A.1 of \citet{dB01}.
These classifications have been supplemented by the regularly
updated information on the Hyades from WEBDA, in particular the interferometric
survey of Hyades multiplicity by \citet{pa98}. In many
cases, the binary nature of the system may be expected to have little or no
impact on the observed colors; the angular separation of the stars might be too
large for both stars to be observed simultaneously with a standard photometer
or the mass, temperature, and luminosity ratio between the stars
may be too large to affect the colors of the primary star. With a reasonably
extensive photometric sample, we have chosen to be very 
conservative in selecting
stars to define the mean relations which can be definitively classed 
as single. Of the 112 members in Table 1, 54 stars have some indication 
of multiplicity and are noted with a ``B'' in column 3, leaving 58 single stars.

Before compiling the final relation, as a precaution in case some binaries
have been included in the derivation due to inadequate information, one
can construct artificial binary systems and estimate what impact they
might have on the two-color relation. Use has been made of the 
final relation in
constructing the composite pairs by choosing a star of a given color on 
the main sequence, then sequentially combining the fluxes in each
bandpass for a series of stars at decreasing mass/luminosity
along the main sequence. The locations of the composite systems in
the CMD are illustrated in Fig. 1a, while the composite colors produce
the pattern shown in Fig. 1b. In Fig. 1a, the curves at each location 
along the CMD
start with two stars of identical color, then illustrate
the effect of altering the composite system using a secondary which resembles
the next sequential point down on the main sequence. The trend in the 
CMD is well-known and essentially
the same for all photometric systems. Composites of identical stars are located
at the same color as the single stars, but 0.75 mag above the main sequence.
As the secondary star is replaced with a cooler, less luminous star farther
down the main sequence, the color difference initially shifts the 
composite increasingly to the red but at a smaller distance above the
main sequence. Eventually, the decreasing luminosity of the secondary
creates a smaller impact on the primary system; the color and absolute
magnitude of the composite move closer to that of the primary star alone.

For the two-color diagram, the pattern of changes is slightly more
complex because of the changing slope of the mean relation. For the stars
hotter than the sun ($b-y \leq 0.4$), the composite systems are always redder
in $b-y$ than the single stars, with the size of the shift larger for
composites
with the smaller difference in luminosity. The combined change in both indices
moves the stars approximately parallel to the mean relation, implying that
binaries should not be a significant source of scatter among hotter dwarfs
in the $hk, b-y$ relation. The steepening slope for the late G through K
dwarfs adds a twist to this pattern comparable to what is seen in the CMD.
The composite system is always redder than the single primary, but the 
pattern in $b-y$ is to grow redder initially as the secondary is shifted down
the main sequence, hit a maximum difference, then shift blueward toward the
color of the primary star. Moreover, the combined effect of changing $b-y$
and $hk$ places the composite star increasingly above the mean relation
before finally approaching the values of the single primary.  
We conclude then that if composite systems are present within the sample,
they should have little to no impact among the hotter stars. Among the
K dwarfs, companions should shift the system above 
the mean relation, {\it i.e.},
toward smaller $hk$, with the size of the shift depending upon the
exact colors of the two stars. Maximal shifts between 0.05 and 0.10 mag
are expected in the $b-y$ range of 0.45 to 0.65.

Using the 58 single Hyades members, a preliminary two-color relation was derived
by sorting the sample in $b-y$ and, starting at the blue end, 
constructing a five-point, weighted average which then shifted redward by
one point before repeating the process. This was continued until a running 
average was available over the entire color range of the sample. This mean
relation was checked for deviant points whose inclusion produced significant
changes in slope over small changes in $b-y$; one point was removed.
The averaging process was repeated, producing a mean relation which
showed a smooth and
continuous change in slope over the entire color range. Finally, residuals
were calculated by comparing the data points to the mean relation and
tested to ensure that the scatter about the mean curve averaged 0.0
with no color dependence. The final Hyades $hk, b-y$ relation is listed in
Table 2 and shown as the solid line in Fig. 2. The open circles of Fig. 2
are the 57 single, non-deviant stars used to 
define the relation, while the crosses 
are the stars tagged as possible binaries. The two weakest parts of
the relation are the extreme ends of the color range. For $b-y$ below 
0.27 and greater than 0.7, the number of points available to define the
curve is small and may be affected by possible binaries. At the cool end, 
the small numbers were set by the apparent faintness of the stars while at
the blue end, hotter stars were too bright ($V \leq 5.5$) for even the 
smallest telescope accessible to this project. In both of these color ranges,
it has been assumed that the Hyades relation has the same relative shape as
that found for the field stars of comparable color. The Hyades members have
been used essentially as a means of fixing the location of the curve
in these regions, not defining the shape of the curve. In short, the
extreme ends of the relation should only be used with caution.

Excluding the one deviant star, the 57 single stars scatter about 
the mean two-color relation with average residuals in $hk$, in the sense 
(OBS-REL), of $+0.001 \pm 0.019$ mag (s.d.).  If binary stars are considered
together with single stars, and if two additional deviant stars are
excluded, the remaining 109 stars show average
residuals of $-0.003 \pm 0.022$, indicating, as predicted, that binarity
moves a star toward lower observed $hk$ relative to the mean relation.
The three stars which exhibit the largest deviation from the mean relation
are the binary stars HIC 19591 and VB/WEBDA 185 (filled squares in Fig. 2), 
while the deviant single star
is HIC 19808 (HG7-135; filled circle in Fig. 2). The 
single star is peculiar in that its colors
imply that the absorption in the region of Ca II H and K is anomalously
large. Since the two-color relation flattens out in the $b-y$ region 
occupied by this star ($\sim$0.7), one cannot construct the indices demonstrated
by this star through combinations of any stars on the standard relation.
Moreover, as will become apparent in the next section, no field star has been
observed which occupies a comparable region of the $hk, b-y$ diagram.
Since one would expect that most forms of stellar interaction would enhance
chromospheric emission leading to line filling, the source of the anomaly
remains a mystery. The simplest solution may be that the observations are
in error, though the star was observed twice and the agreement between the
observations was quite good.

In contrast, the position of HIC 19591 (BD $+23 \arcdeg$ 635), over 0.3
mag above the mean relation is readily explained. It is known to be a
short-period, tidally-locked triple system that exhibits high levels of
chromospheric activity \citep{GG81, BO86} and an anomalous Li abundance
\citep{TH93}. VB/WEBDA 185 is a composite composed of a K3 and a K7 dwarf,
separated in absolute magnitude by only 1.6 mag \citep{GG81}. The system
does not show significant evidence for anomalous chromospheric emission
\citep{st91}. The deviation from the mean relation reflects the optimum
difference in color and intrinsic luminosity required to shift the location of
the system relative to that of the mean relation. The intrinsic colors of
the stars place them in the ideal $b-y$ regime for shifts between 0.05 and
0.10 mag in $hk$ relative to the mean relation, as observed.

\section{The Field Star Comparison} 
With the Hyades $hk, b-y$ relation established, the next task is to determine
what, if anything, distinguishes the Hyades relation from the typical field star
in the solar neighborhood. As stated earlier, expectations are that
the Hyades will establish an approximate lower bound for hotter field stars
due to its high [Fe/H], but remain typical of most stars at lower temperatures
due to the saturation of the Ca II lines. To ensure that any anomalies
between the Hyades and the field stars are not a product of zero-point offsets
within the photometry, one additional check has been made.
In Sec. 2 it was demonstrated that our merged $V$ and $b-y$ data are 
consistent
with the system of \citet{OLS93}. If we compare
the $hk,b-y$ indices for Hyades stars to field dwarfs from the \citet{TAT95}
catalog, can we be certain that the dwarfs
in that catalog are indeed on a color system consistent with \citet{OLS93}?

A set of 401 dwarfs was compiled representing
the overlap between the catalogs of \cite{OLS93} and \citet{TAT95}. 
The stars cover a brightness range
from $V = 4.8$ to $V = 11.6$, and a color range from $(b-y) = 0.08$ to
1.1;  the great majority of these stars have colors between $(b-y)$ of
0.2 and 0.7.  No giants were included in this overlap set, and a number
of stars with discrepant $V$ magnitudes were excluded.
The mean differences in the sense (OLS93 - ATT95) are $0.0007 \pm 0.0006$
(s.e.m.)
for $V$ and $-0.0007 \pm 0.0003$ (s.e.m.) for $(b-y)$.  There are 
no discernible
trends with respect to color or magnitude for $\Delta V$ or $\Delta (b-y)$,
confirming the comparison made during the construction of the original
catalog \citep{TAT95}.

\subsection{The Hotter Dwarfs}
To illustrate the trends in the two-color diagram at reasonable resolution, we
divided the sample in two color ranges with the break point
at $b-y$ = 0.45. For the hotter stars not classified as giants, Fig. 3 shows
the field star sample from \citet{TAT95} with the Hyades mean 
relation superposed as a solid line.
Over the F-star spectral range, $b-y$ between 0.2 and 0.4, the mean relation
defines an approximate lower bound to the field star sample. This is 
expected if the $hk$ index is predominantly a metallicity indicator
and the scatter of stars below the Hyades is, in fact, composed mostly
of stars more metal rich than the Hyades. The few stars that fall well below
the Hyades relation and below virtually all the field stars are supergiants and
not relevant to the discussion.

We wished to probe the metallicity sensitivity of the $hk$ index
as well as investigate the possibility that it might saturate, or even
reverse, as a metallicity indicator for [Fe/H] values greater than the 
Hyades.  One relatively straight-forward way to do this is to construct
an index, $\delta hk$, that describes the difference, in the sense (Hyades
 - star), of a star's $hk$ index from the Hyades $hk$, $b-y$ sequence, and 
then compare $\delta hk$ to photometric or spectroscopic determinations 
of [Fe/H].  

Overlap of the $hk$ catalog \citep{TAT95} with catalogs of high dispersion
spectroscopic abundances is not tremendously large, consisting of 
50 non-binary stars from the catalogs of \citet{edv93} and \citet{ch20} that
fall within the $b-y$ limits of 0.22 and 0.38, a color range for which
we may also utilize the photometric metallicity calibration for F stars
of \citet{SN89}. \citet{ch20} demonstrate effectively that the [Fe/H] values 
in these two spectroscopic catalogs are mutually consistent;  Fig. 4 
illustrates the relationship between [Fe/H]$_{SPEC}$
and $\delta hk$ for 50 F dwarfs.  The form of the relationship 
is [Fe/H]$ = -5.60 \delta hk + 0.065$, where the intercept implies the
[Fe/H] value for $\delta hk = 0.0$, {\it i.e.}, Hyades abundance.  The
dispersion of points about the mean line, 0.08 dex, makes this result
formally consistent with a presumed abundance for the Hyades of
[Fe/H] = 0.125.

We compared the performance of $\delta hk$ relative to photometric
metallicity indicators as well, a considerably less straightforward task but
one that recovers the advantages of several large Str\"omgren catalogs.  The
F star calibration of \citet{SN89} is ideal for this temperature range and is
constrained by limits on applicable $b-y$, $m_1$ and $c_1$.   A few features
deserve note:  several Hyades stars were included in the development of the
calibration, but it was not primarily intended for use at near-solar abundances,
an issue that can be addressed by small additive corrections to the [Fe/H] scale.
Small differences that exist between large catalogs on the Str\"omgren system
can lead to slightly different results as well.  The \citet{SN89} calibration
is consistent with their photometry;  other catalogs may be transformed
to their catalog system using published precepts.  For example, 37 Hyades
F dwarfs are included in the photometric catalog of \citet{OLS83}.  If the
abundance calibration of \citet{SN89} is used, an abundance [Fe/H] $ =
-0.01 \pm 0.09$ is obtained.  A slightly different photometric catalog,
\citet{OLS93} also contains 20 Hyades F dwarfs.  The photometric calibration
provides an estimate of [Fe/H] $= -0.06 \pm 0.07$ for these indices. This
estimate is raised if the Str\"omgren indices of the \citet{OLS93} catalog
are mapped to the system of the calibration, \citet{SN89}.  These precepts
are provided by \citet{OLS93} and produce an estimate of [Fe/H] for the
Hyades stars of $0.04 \pm 0.09$, about 0.1 dex higher than untransformed
indices imply. Thus, slight differences between photometric
scales can lead to modest differences in calculated abundances which may
be adjusted by zero point corrections.
There are also concerns about the {\bf scale} of the photometric estimators of
[Fe/H] for F stars.  \citet{AL96} suggest an adjustment to abundances determined
using the \citet{SN89} F star calibration of the form:  [Fe/H]$_{SPEC} =
1.22 [Fe/H]_{PHOT} + 0.125$, where the intercept is fixed to ensure a
Hyades abundance of $+0.125$. 

With these points in mind, we constructed a comparison of 
$\delta hk$ to photometric abundances, using the two largest, high-precision, 
F-star catalogs on the $uvby$ system with the greatest overlap
with the $hk$ catalog \citep{TAT95}, those of \citet{GR76} for the
brightest stars and \citet{OLS83} for stars at fainter magnitudes. For 
each star between $b-y$ = 0.22 and 0.38 common to the $uvby$ and $hk$
systems, $\delta hk$ is calculated while $uvby$ data for each star 
was processed through the F-star metallicity 
calibration of \citet{SN89}, using the limits on the range of indices for
which the calibration is applicable to further restrict the sample. 
If the indices from \citet{GR76} and \citet{OLS83} are mutually 
consistent, stars of Hyades abundance should have $\delta hk$ values
of 0.0.  Without the adjustment to slope or zero point, the photometric
calibration for Hyades stars yielded a value of [Fe/H] = 0.0, which
is increased to 0.125 by the linear adjusted scale.
 
With these assumptions in mind, Fig. 5 provides a comparison of
photometric estimates of [Fe/H] to $\delta hk$ values for 324 stars.
The circles represent stars from \citet{OLS83}, the crosses data
from \citet{GR76}.   A number of conclusions can be derived from
this diagram.  First, there is no inconsistency of slope or zero point
to indicate that the two photometric samples are anything other than
mutually consistent.  As hoped, at $\delta hk = 0.0$, {\it i.e.}, Hyades
abundance by definition, the implied [Fe/H] value lies between
+0.05 and +0.15, demonstrating that the Hyades metallicity 
scale is consistent with that defined by the field stars. A line
fitted through the points has a slope of -5.8, an intercept of
+0.10, and a dispersion of $\pm 0.12$ dex, expected from photometric
errors alone and in spite of no applied corrections for reddening or
any inhomogeneities within the samples of stars.  The value of the
slope is close enough to that derived from data presented in
Fig. 4 to validate the amplification of the photometric abundance
scale advocated by \citet{AL96}.
It is concluded that for F dwarfs, the Hyades mean relation defines a 
reliable isometallicity line at [Fe/H] = +0.125. 
    
\subsection{The Cooler Dwarfs}
Fig. 6 shows the unevolved cooler field stars from \citet{TAT95} relative to the 
Hyades mean relation. Surprisingly, the Hyades relation crosses over the 
color-color band defined by the field stars and becomes an approximate upper
bound to the sample, rather than a lower limit. This pattern extends from
$b-y$ $\sim$ 0.5 to 0.65, where the flattening of the Hyades relation, coupled with
the small number of points defining the mean relation, makes it difficult to
pin down exactly where the Hyades is located relative to the field stars.

If the peculiar location of the Hyades relation is not the product
of photometric error, what might be its cause?  Three possibilities come to mind:

1){\it The location of the two-color relation is linked to the {\it Hyades anomaly}}.
The Hyades anomaly refers to observation that the unevolved F stars 
in the Hyades have $c_1$ indices that place them above the unevolved 
field stars at the
same temperature \citep{CB69}. Since $c_1$ is an indicator of surface 
gravity through the Balmer discontinuity, the higher
$c_1$ values at a given $b-y$ or H$\beta$ imply that the F dwarfs in the 
Hyades are more luminous/evolved than the field stars of comparable age.
Over the last 15 years, numerous solutions have been proposed and questioned,
including filter errors \citep{EG94,JT95}, starspots triggered by
chromospheric activity \citep{RO84, SO89}, some form of abundance anomaly, 
including overall metallicity \citep{BA74, AL86}, CN variations 
\citep{EG82, BT83}, 
and helium \citep{ST82, AL86}. Others have questioned the reality of the phenomenon,
though the work of \citet{DO90} indicates that something that is not
predicted by simple, standard models is happening. Finally, \citet{NI88} 
would appear to confirm that the Hyades cluster is
not alone in this peculiarity, while demonstrating that the anomaly is not
a simple age or metallicity effect.

There are two reasons to discount the relevance of the Hyades anomaly in the
current discussion. First, if a star has a higher $c_1$ index at a given
color, this indicates that the star is fainter than expected in the ultraviolet
region of the spectrum. The cooler Hyades dwarfs look metal-deficient, {\it i.e.}, their
energy distribution is skewed to higher ultraviolet flux, though one could 
claim that over the narrow bandpass of the $Ca$ filter alone, the flux is higher
than expected. Second, photoelectric observations of the cooler stars by
\citet{rf92} demonstrate that the Hyades dwarfs have $c_1$ indices consistent
with the mean relation for field stars and show no evidence for the anomaly
identified among the hotter stars.

2) {\it The stars in the Hyades are younger than the average field star, so the 
increased chromospheric activity compensates for the higher metallicity by
filling in the Ca II H and K lines and making the star appear metal-deficient.}
If correct, stars that lie above the Hyades relation in the two-color 
diagram must be metal-poor and/or chromospherically active, the latter due
to a younger age or environmentally-induced activity. The fact that the
observed stars known to be chromospherically active to an anomalous degree
sit well above the Hyades relation is consistent with this option. 

3) {\it Though the metallicity effect on the Ca II lines should increase
the value of $hk$ at a given $b-y$ as [Fe/H] increases, above a 
critical value of [Fe/H] the index hits a maximum and then decreases
with increasing [Fe/H].} Among the most metal-rich stars in the
solar neighborhood, Hyades members are positioned {\bf above} stars of lower
metallicity in the two-color diagram. Therefore, stars even more metal
rich than the Hyades should fall above the mean relation in Fig. 6.

To sort through the possibilities, we have identified 59 stars that lie
above the Hyades mean relation between $b-y$ = 0.50 and 0.8. As a first
step, the $Hipparcos$ catalog was searched and all but 5 stars had a measured
parallax. Of the remaining 54, all but 8 had $\sigma$/$\pi$ of 0.16 or less, 
implying that reliable absolute magnitudes were possible. The CMD for the
remaining 46 stars is shown in Fig. 7 where the error bars drawn are based
upon the one sigma uncertainty in the parallax. What is immediately apparent
is that the sample contains a number of stars that should legitimately be
classified as subgiants/giants, {\it i.e.}, any star with $M_V$ below 5.0. To
this sample of 8 stars can be added 3 stars that have measured parallaxes
with reasonable errors, but absolute parallaxes so small the stars must be
at large distance and therefore must be intrinsically more luminous than a
dwarf. The misclassification of these stars as dwarfs explains their
anomalous location in the two-color diagram because the mean relation for
solar-metallicity giants is shifted approximately 0.1 mag above that for the
dwarfs due to surface gravity effects. Though these stars will be excluded from
further discussion, we note that two of the stars are of interest because they
are anomalous even for giants. HD 13435 is the reddest star
at the base of the giant branch, occupying a position that is undoubtedly
heavily dependent upon its exceptionally high [Fe/H] \citep{fa97}. HD 89499
sits well above the mean relation due to classification as both a subgiant
and a chromospherically-active, short-period, tidally-locked binary 
\citep{rd95}.

The next step in sorting through the remaining 48 stars is to estimate their
metallicity and determine if they fall into either extreme end of the abundance
range or cover the entire spread. As one might expect, the information 
available for the stars is mixed. The spectroscopic abundance catalog of
\citet{CA01} includes high dispersion results for 10 of the stars, 11 if one
includes the value for HD 61606A as representative of HD 61606B. For stars
with multiple abundance estimates, an unweighted average has been taken.
Beyond spectroscopy, we can appeal to photometry for abundances. A survey
of the $uvby$ catalog of \citet{HM98} identifies 18 additional stars with
a full set of $uvby$ indices, of which 11 fall within the calibration limits
of the G-star calibration of \citet{SN89}. For stars with [Fe/H] below --0.4,
the abundances derived from the calibration were adopted as is without any
adjustment to the zero-points of the photometry or the metallicity scale.
For stars with [Fe/H] above this limit, the photometry was also uncorrected
for any zero-point differences with \citet{SN88}, but the photometric
abundances were derived by comparison to the photometric abundance derived
for Hyades stars at the same color using the \citet{SN89} calibration. 
The reason for this change is discussed in detail in \citet{TW02} and
arises from a flaw in the photometric abundance calibration that leads to
a severe underestimate of [Fe/H] for cooler G and early K dwarfs
at solar abundance and above. For derivation of the final abundances, it
has been assumed that [Fe/H] for the Hyades is +0.125. Finally, 
\citet{AL96} have derived a photometric abundance of --1.5 for HD 4967,
while \citet{LCL} obtain +0.32 for HD 95741. 

Of the 24 stars with some semblance of an abundance estimate, 9 were below
[Fe/H] = --0.70, implying that their position in the two-color diagram is the
product of low [Fe/H]. Of the remaining 15 stars, none were found between
[Fe/H] = --0.20 and --0.70. The abundance distribution for these stars is
plotted in Fig. 8 and clearly shows that, despite the lack of homogeneity
in the sample abundances, the stars are heavily weighted toward Hyades
abundance or higher, as predicted if the shift of the mean relation is due
to a metallicity effect.  All stars with [Fe/H] below --0.70 are contained 
in the lowest bin.

The final classification property to check is that of chromospheric activity.
Seven of the stars are notable for their unusual degree of chromospheric
activity, in many cases leading to classification as $BY Dra$ stars. HD
89499 is a subgiant and has been noted earlier. BD $-0 \arcdeg$ 4234 
is a metal-poor
dwarf in a tidally-locked binary system \citep{rd95} 
and was included in the metal-poor
classification above. Thus, the presence of 5 additional stars above the
Hyades in the two-color relation can now be clarified, leaving 19 with
insufficient information to determine their status. An attempt has been
made to discover if there is any general trend with chromospheric
activity beyond the obvious effect for stars of extreme activity.
The stellar sample with measurements of the chromospheric index, 
$R'_{HD}$
compiled by \citet{RP00} has been matched with the $Caby$ catalog, producing
an overlap of 189 stars. Unfortunately, only 14 stars lie 
within the critical color
range redder than $b-y$ = 0.50. For each of these stars, a residual 
$\delta R'_{HK}$ was calculated by subtracting the observed index from a
linear lower limit drawn through the sample as a function of color. No
correlation was found between this chromospheric index and the position of the
star relative to the Hyades in the $hk, b-y$ diagram at the same color though,
again, the sample is small.

We note that in some
cases, stars have been categorized in a way that might indicate anomalously
high [Fe/H], {\it e.g.}, inclusion in a metal-rich moving group, but such a
designation was not deemed reliable enough to justify the assumption of
high [Fe/H]. Moreover, among the coolest dwarfs, many were selected because
of their specification as being within such a group, making the sample less
than random.

In Fig. 9, we repeat Fig. 7 with the error bars removed. Stars classed as
metal-poor have been removed. Filled circles identify all stars with [Fe/H]
above --0.20, triangles identify stars that are chromospherically active,
open circles are stars with no definitive information, and the crosses 
show the location of stars within the Hyades as compiled in \citet{TW02}.
In the range from $b-y$ = 0.50 to 0.65, the majority of the sample is tagged
as metal-rich and occupies a position in the CMD consistent with that
designation, {\it i.e.}, on or above the Hyades main sequence. Three stars
bluer than $b-y$ = 0.66 that occupy a position consistent with high
[Fe/H] but have no spectroscopic abundance determinations are HD 57095,
HD 76378, and HD 219495.

Among the stars classed as chromospherically active, there is no apparent
trend relative to the main sequence, with stars scattering above and below
the Hyades, as expected if this sample originates from a population with
a broader range in [Fe/H]. The majority of stars for which no
information is available fall in the redder half of the distribution where
spectroscopic data are few and far between; these stars scatter both
above and below the Hyades. While it is possible that these stars represent
a mixture of both metal-rich and chromospherically-active stars, we 
emphasize again the uncertainty in the definition of the mean Hyades relation
among the coolest dwarfs. It may well be that with a larger array of cool Hyades
and field star photometry, the position of the mean relation will shift or
the samples for all metallicities will converge to a common relation
insensitive to [Fe/H], as originally expected for the K dwarfs.

\section{Summary and Conclusions}                                     
An extensive sample of $Caby$ photometry of Hyades dwarfs from A through early M
has been compiled and analyzed. For single stars and simple, composite, binary 
systems, the mean relation is well-defined over the color range from $b-y$
= 0.25 to 0.65. For the hotter portion of the color range, the effect of
binarity on the two-color diagram is minimal. Among cooler dwarfs, the presence
of a fainter, secondary star tends to shift the system above the mean relation,
{\it i.e.}, to lower $hk$ at a given $b-y$, simulating 
a lower [Fe/H], with a maximum offset between 0.05 and 0.10 mag. Stars with
extreme degrees of chromospheric activity, particularly $BY Dra$ stars, appear
anomalously metal-poor due to line-filling by emission within the Ca II H 
and K lines.

Among the F dwarfs, the Hyades relation defines a reliable lower bound to the
distribution of field stars for a fixed [Fe/H] of +0.125. By comparison 
to the spectroscopic metallicity scale of \citet{edv93}, 
it is found that the $hk$ index is linearly correlated with
[Fe/H] over the range from +0.5 to --1.0, with almost twice the sensitivity
to abundance changes compared to $m_1$. A simple test of the utility of $hk$
at super-metal-rich levels is provided by the sample of stars recently
identified as having planets as compiled by the University of California
Planet Search Team (http://exoplanets.org). Of the 76 systems listed, 16
are in the $hk$ Catalog \citep{TAT95}. Of these, 6 are within the
F-star color limits and have [Fe/H] between +0.17 and +0.34, with a mean
of $+0.23 \pm 0.08$. Line saturation does not appear to be a problem for
the hotter stars. 

In sharp contrast, as one extends the Hyades data toward 
cooler temperatures, the 
mean two-color relation crosses the distribution of nearby field
stars, producing an approximate upper bound to the sample for stars in the
color range from $b-y$ = 0.50 to 0.65. From an analysis of the stars that
lie even higher than this relation in the $hk, b-y$ diagram, the primary
source of the effect appears to be the high metallicity of the Hyades cluster.
The implication is that the index does saturate at a given color for
stars near solar abundance, but additional increases in [Fe/H] lower the index,
placing super-metal-rich stars in the same region of
the diagram as stars with [Fe/H] near --0.7 or lower. 

An alternative explanation
may be provided by the synthetic indices developed from model
atmospheres by \citet{SO93}. While the models imply that $hk$ should remain
metallicity sensitive to [Fe/H] = +0.5 for all colors, a pattern almost
identical to that found for $hk$ is generated for $C_{RV}$, the photospheric
narrow-band index constructed from the data collected for the Mt. Wilson
survey of stellar chromospheric emission \citep{VA78}. 
\citet{SO93} explain the crossover of $C_{RV}$ at cooler
temperatures as primarily a result of the metallicity sensitivity of $b-y$,
an effect that disappears with the use of $V-K$. Why this trend exists for
an index based upon a filter over four times wider than those used in
$C_{RV}$ remains a mystery, but it may be more of an indication of the model
shortcomings than of a serious problem with the observed indices. In any
case, the $hk$ index for cooler stars clearly succeeds in achieving the goal
for which it was designed, the identification and calibration of stars
of intermediate to extremely low [Fe/H]. This investigation adds two
options to the list of uses, identification of stars with high levels of
chromospheric activity, as expected, and the totally unexpected
ability to isolate stars of Hyades abundance and higher. Three stars likely
to occupy the last category and deserving of closer study are HD 57095,
HD 76378, and HD 219495. 

The question of the impact of modest, age-dependent
variations in chromospheric emission remains unresolved, though indications
from the very limited sample available to date are that the effect on $hk$ is
small to negligible. Observations of the G and K dwarfs in nearby open 
clusters of
comparable age but lower [Fe/H] than the Hyades could help to resolve this
issue.

\acknowledgements
It is a pleasure to thank Dr. Rocha-Pinto for supplying us with a copy
of the sample used in the study of the age-metallicity-relation from
chromospheric emission.  Although our thanks must reach back many years,
we are indebted to the staffs of both NOAO observatories.
We made extensive use of the WEBDA and
SIMBAD databases, and are happy to acknowledge our appreciation to
their creators and contributors.  We are also indebted to the
very helpful comments and constructive criticisms of the referee.

\clearpage
\figcaption[fig1.eps]{The effects of binarity 
on the $hk$, $b-y$ observations of cluster stars. (a) The CMD for 
stars combined with companions of decreasing
temperature on the main sequence. (b) The $hk$, $b-y$ diagram for stars combined
with companions of decreasing temperature on the main sequence. \label{fig1}}

\figcaption[fig2.eps]{The $hk, b-y$ diagram for the Hyades members. Open
circles are single stars, crosses are probable multiple systems, and filled
symbols are anomalous points excluded from the derivation of the mean
relation, given by the solid curve. \label{fig2}}

\figcaption[fig3.eps]{$hk, b-y$ data for hotter field stars with the Hyades
relation superposed. \label{fig3}}

\figcaption[fig4.eps]{Comparison of spectroscopic abundances from the catalogs 
of \citet{edv93} and \citet{ch20}, to the metallicity index, $\delta hk$. 
Solid line is a linear fit through the points. 
\label{fig4}}

\figcaption[fig5.eps]{Comparison of photometric abundances from the $uvby$  photometry
of \citet{GR76} (crosses) and \citet{OLS83}(circles) to the metallicity
index, $\delta$$hk$. \label{fig5}}  

\figcaption[fig6.eps]{Same as Fig. 3 for the cooler field dwarfs. \label{fig6}}

\figcaption[fig7.eps]{CMD of stars located above the Hyades in the $hk, b-y$
diagram using $Hipparcos$ parallaxes. Error bars show the uncertainty
based upon the one sigma error in the parallax. \label{fig7}}

\figcaption[fig8.eps]{Abundance distribution for stars above the Hyades
relation. Abundances represent a mixture of spectroscopic and photometric
data. All stars with [Fe/H] below --0.70 are contained in the lowest bin. \label{fig8}}

\figcaption[fig9.eps]{Same as Fig. 7 with evolved stars and metal-poor stars 
removed. Filled circles are the stars with [Fe/H] above --0.2, triangles
are stars that are chromospherically active, open stars are data points
without any clarifying information, and crosses are observed single
stars within the Hyades. \label{fig9}} 

\begin{thebibliography}{}
\bibitem[Alexander (1986)]{AL86} Alexander, J. B. 1986, \mnras, 220, 473
\bibitem[Alonso et al. (1996)]{AL96} Alonso, A., Arribas, S., \& Martinez-Rogers,
C. 1996, \aaps, 117, 227
\bibitem[Anthony-Twarog et al. (1991)]{ATT91} Anthony-Twarog, B. J., Laird, J. B., Payne, D., 
\& Twarog, B. A. 1991, \aj, 101, 1902
\bibitem[Anthony-Twarog et al. (2000)]{ATT00} Anthony-Twarog, B. J., Sarajedini, A., Twarog, 
B. A., \& Beers, T. C. 2000, \aj, 119, 2882
\bibitem[Anthony-Twarog et al. (1992)]{AST92} Anthony-Twarog, B. J., Shawl, S.
J., \& Twarog, B. A. 1992, \aj, 104, 2229
\bibitem[Anthony-Twarog \& Twarog (1998)]{ATT98} Anthony-Twarog, B. J., \& Twarog, B. A., 
1998, \aj, 116, 1902
\bibitem[Anthony-Twarog et al. (1995)]{ATC95} Anthony-Twarog, B. J., Twarog, B. A., \& Craig, 
J. 1995, \pasp, 107, 32
\bibitem[Baird (1996)]{BD96} Baird, S. R. 1996, \aj 112, 2132
\bibitem[Barry (1974)]{BA74} Barry, D. C. 1974, \aj, 79, 616
\bibitem[Bopp et al. (1986)]{BO86} Bopp, B. W., Africano, J. L., \& Goodrich,
B. D. 1986, \pasp, 98, 457
\bibitem[Brown \& Twarog (1983)]{BT83} Brown, J. A., 1983, \aj, 88, 678
\bibitem[Carney (1979)]{CA79} Carney, B. W. 1979, \apj, 233, 211
\bibitem[Cayrel de Strobel et al. (2001)]{CA01} Cayrel de Strobel, G., Soubiran,
C., \& Ralite, N. 2001, \aap, 373, 159
\bibitem[Chen et al. (2000)]{ch20} Chen, Y.Q. et al., 2000, \aaps 141, 491
\bibitem[Crawford (1975)]{CR75} Crawford, D. L. 1975, \aj, 80, 955
\bibitem[Crawford \& Barnes (1969)]{CB69} Crawford, D. L., \& Barnes, J. V. 1969,
\aj, 74, 407
\bibitem[Crawford \& Perry (1966)]{cp66} Crawford, D. L. \& Perry, C. L. 1966, \aj, 71, 206
\bibitem[de Bruijne et al. (2001)] {dB01} de Bruijne, J. H. J., Hoogerwerf,
R., \& de Zeeuw, P. T. 2001, \aap, 367, 111                     
\bibitem[Deming et al. (1977)]{DEM77} Deming, D., Olson, E. C., \& Yoss, K. M.  1977, \aap, 
57, 417
\bibitem[Dobson (1990)]{DO90} Dobson, A. K. 1990, \pasp, 102, 88
\bibitem[Edvardsson et al. (1993)]{edv93} Edvardsson,  et al., 1993, \aaps,  275, 101
\bibitem[Eggen (1982)]{EG82} Eggen, O. J. 1982, \apjs, 50, 22
\bibitem[Eggen (1994)]{EG94} Eggen, O. J. 1994, \aj, 107, 594
\bibitem[Favata et al. (1997)]{fa97} Favata, F., Micela, G., \& Sciortino, S.
1997, \aap, 323, 809
\bibitem[Griffin et al. (1988)]{GR88} Griffin, R. F., Gunn, J. E., Zimmerman, B. A., \& 
Griffin, R. E. M. 1988, \aj, 96, 172
\bibitem[Griffin \& Gunn (1981)]{GG81} Griffin, R. F., \& Gunn, J. E. 1981,
\aj, 86, 588
\bibitem[Gr$\o$nbech \& Olsen (1976)]{GR76} Gr$\o$nbech, B., \& Olsen, E. H. 1976,
\aaps, 25, 213
\bibitem[Hanson \& Vasilevskis (1983)]{HA81} Hanson, R. B. \& Vasilevskis, S. 1983, \aj, 88, 
844
\bibitem[Hauck \& Mermilliod (1998)]{HM98} Hauck, B., \& Mermilliod, M. 1998,
\aaps, 129, 431
\bibitem[Hintz et al. (1998)]{HI98} Hintz, M. L., Joner, M. D., \& Hintz, E. G.  1998, \aj, 
116, 2993
\bibitem[Joner \& Taylor (1995)]{JT95} Joner, M. D., \& Taylor, B. J. 1995,
\pasp, 107, 124
\bibitem[Laird et al. (1988)]{LCL} Laird, J. B., Carney, B. W., \& Latham, D. W. 
1988, \aj, 95, 1843
\bibitem[Meusinger et al. (1991)]{ME91} Meusinger, H., Reimann, H. -C., \& Stecklum, B. 1991, 
\aap, 245, 57
\bibitem[Nissen (1988)]{NI88} Nissen, P. E. 1988, \aap, 199, 146
\bibitem[Olsen (1983)]{OLS83} Olsen, E. H. 1983, \aaps, 54, 55
\bibitem[Olsen (1993)]{OLS93} Olsen, E. H. 1993, \aaps, 102, 89
\bibitem[Olsen (1994)]{OLS94} Olsen, E. H. 1994, \aaps, 106, 257
\bibitem[Patience et al. (1998)]{pa98} Patience, J., Ghez, A. M., Reid, I. N.,
Weinberger, A. J., \& Matthews, K. 1998, \aj, 115, 1972
\bibitem[Perry (1969)]{p69} Perry, C. L. 1969, \aj, 74. 705
\bibitem[Perryman et al. (1998)]{PE98} Perryman, M. A. C., Brown, A. G. A., 
Lebreton, Y., Gomez, A., Turon, C., Cayrel de Strobel, G., Mermilliod, J. -C.,
Robichon, N., Kovalevsky, J., \& Crifo, F. 1998, \aap, 331, 81 
\bibitem[Reglero et al. (1992)]{rf92} Reglero, V., Fabregat, J., Suso, J. 1992, \aaps, 92, 565
\bibitem[Rey et al. (2000)]{RE00} Rey, S. -C., Lee, Y. -W., Joo, J. -M., Walker, A., \& Baird, 
S. 2000, \aj, 119, 1824
\bibitem[Rocha-Pinto et al. (2000)]{RP00} Rocha-Pinto, H. J., Maciel, W. J., Scalo, 
J., \& Flynn, C. 2000, \aap, 358, 850 
\bibitem[Rose (1984)]{RO84} Rose, J. A. 1984, \aj, 89, 1238
\bibitem[Ryan \& Deliyannis (1995)]{rd95} Ryan, S. G., \& Deliyannis, C. P. 1995,
\apj, 453, 819
\bibitem[Schuster \& Nissen (1988)]{SN88} Schuster, W. J., \& Nissen, P. E. 
1988, \aaps, 73, 225
\bibitem[Schuster \& Nissen (1989)]{SN89} Schuster, W. J., \& Nissen, P. E. 1989, \aap, 221, 
65
\bibitem[Soderblom (1989)]{SO89} Soderblom, D. R. 1989, \apj, 342, 823
\bibitem[Soon et al. (1993)]{SO93} Soon, W. H., Zhang, Q., Baliunas, S. L., Kurucz, R. L. 
1993, \apj, 416, 787
\bibitem[Sowell \& Wilson (1993)]{sw93} Sowell, J.R. \& Wilson, J. W. 1993, \pasp, 105, 36
\bibitem[Stauffer et al. (1991)]{st91} Stauffer, J. R., Giampapa, M. S., 
Herbst, W., Vincent, J. M., Hartmann, L. W., \& Stern, R. A. 1991, \apj,
374, 142
\bibitem[Str\"omgren et al. (1982)]{ST82} Str\"omgren, B., Gustafsson, B., \&
Olsen, E. H. 1982, \pasp, 94, 5
\bibitem[Taylor \& Joner (1992)]{tj92} Taylor, B. J. \& Joner, M. D. 1992, \pasp, 104 911
\bibitem[Thorburn et al. (1993)]{TH93} Thorburn, J. A., Hobbs, L. M., 
Deliyannis, C. P., \& Pinsonneault, M. H. 1993, \apj, 415, 150
\bibitem[Twarog (1980)]{TW80} Twarog, B. A. 1980, \apj, 242, 242
\bibitem[Twarog \& Anthony-Twarog (1995)]{TAT95} Twarog, B. A., \& Anthony-Twarog, B. J. 1995, 
\aj, 109, 2828
\bibitem[Twarog \& Anthony-Twarog (1996)]{TAT96} Twarog, B. A., \& Anthony-Twarog, B. J. 1996, 
\aj, 112, 1500
\bibitem[Twarog et al. (2002)]{TW02} Twarog, B. A., Anthony-Twarog, B. J., \&
Tanner, D. 2002, \aj, submitted 
\bibitem[Twarog et al. (1997)]{TAT97} Twarog, B. A., Ashman, K. M., \& Anthony-Twarog, B. A. 
1997, \aj, 114, 2556  
\bibitem[Van Bueren (1952)]{VB52} Van Bueren, H. G. 1952, Bull. Astr. Inst.  Neth., 11, 385
\bibitem[Vaughan (1980)]{VA80} Vaughan, A. H. 1980, \pasp, 92, 392
\bibitem[Vaughan et al. (1978)]{VA78} Vaughan, A. H., Preston, G. W., \&
Wilson, O. C. 1978, \pasp, 90, 267
\bibitem[Wilson (1963)]{WI63} Wilson, O. C. 1963, \apj, 138, 832

\end{thebibliography}
\enddocument